\documentclass[aps, prb, reprint, superscriptaddress, amsmath, amsfonts, amssymb]{revtex4-1}
\usepackage{dsfont}
\usepackage{graphicx}
\usepackage{extarrows}
\usepackage{color}
\usepackage{feynmp}
\usepackage{ifpdf}
\ifpdf
\fi

\DeclareMathOperator{\sgn}{sgn}

\newcommand{\nn}[0]{ \nonumber \\}

\newcommand{\blr}[1]{ \left( #1 \right) }
\newcommand{\bclr}[1]{ \left[ #1 \right] }
\newcommand{\bslr}[1]{ \left\lbrace #1 \right\rbrace }
\newcommand{\abs}[1]{ \left| #1 \right|}
\newcommand{\Blr}[1]{ \Big( \! #1 \! \Big) }

\newcommand{\ahalf}[0]{ \tfrac{1}{2}}
\newcommand{\aquater}[0]{ \tfrac{1}{4}}
\newcommand{\otp}[1]{ \tfrac{1}{\blr{2 \pi}^{#1}}}

\newcommand{\pri}[0]{^{\prime}}

\renewcommand{\vec}[1]{ \boldsymbol{ #1 } }
\newcommand{\mat}[1]{ \underline{\vec{#1}} }
\renewcommand{\vr}[0]{ \vec{r} }
\newcommand{\vq}[0]{ \vec{q} }
\newcommand{\vk}[0]{ \vec{k} }
\newcommand{\vD}[0]{ \vec{\nabla} }
\newcommand{\vm} {\vec{m}}
\newcommand{\vB} {\vec{B}}

\newcommand{\be}{\begin{equation}}
\newcommand{\ee}{\end{equation}}
\newcommand{\bfsigma}{\vec{\sigma}}

\newcommand{\kf}[0]{k_{\mathrm{F}}}
\newcommand{\kfu}[0]{k_{\mathrm{F}\uparrow}}
\newcommand{\kfd}[0]{k_{\mathrm{F}\downarrow}}
\newcommand{\mub}[0]{\mu_{\mathrm{B}}}

\newcommand{\ind}[1]{ \int\!\!\mathrm{d}#1 \; }
\newcommand{\indotp}[2]{ \int\!\!\tfrac{\mathrm{d}#1}{\blr{2 \pi}^{#2}} \; }

\definecolor{gg}{rgb}{0.0,0.588,0.0}
\definecolor{rr}{rgb}{0.588,0.0,0.0}
\definecolor{bb}{rgb}{0.0,0.0,0.588}

\begin{document}

\title{ Transverse and longitudinal gradients of the spin magnetization \\ in Spin-Density-Functional Theory }

\author{F. G. Eich}
\affiliation{Department of Physics and Astronomy, University of Missouri, Columbia, Missouri 65211, USA} 
\author{S. Pittalis}
\affiliation{CNR -- Istituto Nanoscienze, Centro S3, Modena, Italy}
\author{G. Vignale}
\affiliation{Department of Physics and Astronomy, University of Missouri, Columbia, Missouri 65211, USA} 

\begin{abstract}
 We derive the gradient expansion for the exchange energy of a spin-polarized electron gas by perturbing the uniformly spin polarized state and  thus inducing a small non-collinearity that is slowly varying in space. We show that the exchange-energy contribution due to the induced longitudinal gradient of the spin polarization to the exchange energy differs from the contribution due to the transverse gradient. The difference is present at any non-vanishing spin polarization and becomes larger with increasing spin polarization. We argue that improved generalized gradient approximations of Spin-Density-Functional Theory must account for the difference between the longitudinal and transverse spin stiffness.
\end{abstract}

\date{\today}

\maketitle

\section{ Introduction } \label{intro}

The growing interest in the electronic structure of magnetic materials motivates recent efforts to extend 
the methods of Density-Functional Theory (DFT)\cite{HohenbergKohn:64,KohnSham:65,DreizlerGross:90} to non-collinear spin situations.
For example, the field of spintronics provides exciting potential applications for data storage and manipulation.\cite{ZuticDasSarma:04}
Exciting effects, such as the spin-transfer torque, are already being used in a new generation of random access memories.\cite{RalphStiles:08} An important ingredient of many spintronic devices is spin-orbit coupling.  It provides the mechanism to connect the charge and spin degrees of freedom. Due to this coupling  the spin magnetization exhibits fully its vectorial character, i.e.~, the spin polarization can no longer be viewed as either ``up'' or ``down'',
but its direction may vary in space, giving rise to  \emph{non-collinear}  magnetic structures.  A prominent example of such a structure is the skyrmion~\cite{Skyrme:62}, recently observed in metallic compounds with strong spin-orbit coupling.\cite{Muehlbauer:09,HeinzeBlugel:11}

More than 40 years ago von Barth and Hedin formulated
non-collinear Spin-Density-Functional Theory (SDFT).\cite{BarthHedin:72}
SDFT deals with Hamiltonians for interacting electrons which include a Zeeman
coupling of the spin degrees of freedom to an external magnetic field $\vB(\vr)$. 
Effects of spin-orbit coupling are typically treated perturbatively. In principle they can also be treated non-perturbatively and  self-consistently by
including the spin-currents as additional basic variables,  coupled to external ${\mathrm{SU}\!\blr{2}}$ fields.~\cite{AbedinpourTokatly:10,Bencheikh:03,RohraGoerling:06}
Here, however, we shall only be concerned with the basic formulation of SDFT for non-collinear magnetism.
We remark that consideration of a full-fledged non-collinear SDFT has the advantage of resolving~\cite{Gidopoulos:07} issues
related to non-trivial examples of non-uniqueness between the sets of
external potentials and corresponding densities.~\cite{BarthHedin:72,EschrigPickett:01,KohnUllrich:04,Ullrich:05,CapelleVignale:01}
Resting on this firm basis, we can safely focus on the physics encoded into the exchange-correlation (${xc}$) energy functional.
This functional is the central object that has to be approximated for an actual implementation of SDFT.

The purpose of this paper is to derive a gradient expansion (GEA) for the exchange energy in a non-collinear spin configuration in order to assess the validity of 
existing approximations. We demonstrate through an explicit calculation the
difference between the contributions of longitudinal and transverse gradients of the spin magnetization to the exchange
energy of an electron gas in which the magnitude and direction of the spin polarization vary slowly in space.  The proper treatment of the longitudinal and transverse
spin gradients is of crucial importance for the description of magnetic phase transitions.
Our analysis provides important information for constructing approximate functionals for the ${xc}$ energy in magnetic systems, particularly, if one aims at the description of
extended systems. 

This paper is organized as follows. In Sec.\ (\ref{Functionals}) we review existing approximate functionals for non-collinear  SDFT with a focus on the construction of generalized gradient approximations.  The key results of the paper are surveyed.  Sec.\ (\ref{Gradients}) contains the derivation of the key results, namely the non-collinear gradient expansion for the slowly varying state of the uniform gas at the exact-exchange level. Lastly, Sec.\ (\ref{Discussion}) contains a brief discussion of the prospects for the application of our results to the construction of approximate ${xc}$ functionals for non-collinear spin systems.

\section{Spin stiffness and approximate functionals in SDFT} \label{Functionals}

At the heart of any DFT is the mapping of the interacting system onto an auxiliary non-interacting system,
the so-called Kohn-Sham system. In SDFT one has to solve the following single-particle Pauli equation,
\be\label{KSSDFT}
\left[-\frac{1}{2}
\nabla^2+v_s({\bf r})+\mub \bfsigma 
\vB_s({\bf r})\right]\Phi_i({\bf r})= \epsilon_i \Phi_i({\bf r}) \;,
\ee
where $\Phi_i({\bf r})$ are two-component Pauli spinors. The Kohn-Sham system reproduces
the charge density 
\be\label{pdsdft}
n({\bf r}) = \sum_{i} f_{i} \Phi_i^{\dagger}({\bf r}) \Phi_i({\bf r}) \; ,
\ee
and the spin magnetization
\be
\vm({\bf r}) = - \sum_{i} f_{i} \Phi_i^{\dagger}({\bf r}) \bfsigma 
\Phi_i({\bf r})  \; ,
\ee
of the interacting system, where the ${f_{i}}$ are occupation numbers to be determined self-consistently (often, one may work
within the restriction of integer occupation numbers; i.e. combining a finite number of spin-orbitals into a single Slater determinant).
The effective potentials are decomposed into
\be
v_s({\bf r}) = v({\bf r})+v_H({\bf r})+ v_{xc}({\bf r})\; ,
\ee
and
\be
\vB_s({\bf r}) = \vB({\bf r})+\vB_{xc}({\bf r}) \; ,
\ee
where  $v_H({\bf r})$ is the usual Hartree potential.
The ${xc}$ potentials are functional derivatives of the 
${xc}$ energy $E_{xc}$ with respect to the corresponding 
conjugate densities 
\be
v_{xc}({\bf r}) = \frac{\delta E_{xc}[n,\vm]}{\delta n({\bf r})} \Big|_{\vm}~,
\ee
and
\be
\mub \vB_{xc}({\bf r}) = - \frac{\delta E_{xc}[n,\vm]}{\delta \vm({\bf r})}\Big|_{n} ~.
\ee
$E_{xc}$, a functional of the density ${n\!\blr{\vr}}$ and the spin magnetization ${\vec{m}\!\blr{\vr}}$ in
the case of SDFT, is the key quantity that needs to be approximated.
An important feature of SDFT is the appearance of a \emph{local} torque due to ${xc}$ effects.
This torque enters in the balance equation for the ground-state spin magnetization\cite{CapelleGyoerffy:01}
\be
\vD \cdot \mat{J}_{s}(\vr) + 2 \mub \, \vm(\vr) \times \vB_s(\vr) = 0\;,
\label{spin-balance}
\ee
where the Kohn-Sham spin-current tensor is given by
\begin{align}
  \bigg[ \mat{J}_{s}(\vr) \bigg]_{\alpha \beta} \!\!\!\!\!\!= \tfrac{1}{2 i} 
  \sum_i f_{i}\left\{ \left[ \frac{\partial \Phi_i^{\dagger}(\vr)}{\partial r_{\beta}}  \right] \sigma_{\alpha} \Phi_i(\vr) - \mathrm{h.c.} \right\} ,
\end{align}
and the divergence of this tensor is defined as 
\be
\bigg[\vD \cdot \mat{J}_{s}(\vr)\bigg]_{\alpha} = \sum_{\beta=1}^3 
\frac{\partial}{\partial r_{\beta}} \bigg[\mat{J}_{s}(\vr)
  \bigg]_{\alpha \beta} \; .
\ee
Note that Eq.~(\ref{spin-balance}) admits a non-vanishing \emph{local} torque ${\vm(\vr)\times\vB_{{\rm xc}}(\vr)}$,
even for a vanishing external magnetic field. \emph{Globally} this torque must vanish since the
electron-electron interaction cannot exert a net torque on the system. The so-called zero-torque theorem, 
which follows from the invariance of the ${xc}$ correlation energy under a global rotation of all the spins, reads~\cite{CapelleGyoerffy:01}
\begin{align}
  0 & = \ind{^3r}  \vm(\vr) \times \vB_{xc}(\vr) ~.\label{ZTT}
\end{align}
The presence of a local torque in non-collinear systems has been demonstrated numerically within
an exact-exchange calculation for an unsupported Cr monolayer.\cite{SharmaGross:07} Recent studies
employing semi-local functionals have shown that for this system a local torque is present if the
${xc}$ functional depends on transverse gradients.\cite{BulikScuseria:13,EichGross:13}

The most popular approach to treat non-collinear systems is to use ${xc}$ energy functionals \cite{RappoportBurke:09} borrowed
from collinear SDFT to the non-collinear situation.  
The standard local spin density approximation (LSDA) can be applied straightforwardly to non-collinear system.\cite{KueblerWilliams:88,NordstroemSingh:96,OdaCar:98}
However, the LSDA depends only on the magnitude of the spin magnetization and therefore always yields ${xc}$ magnetic fields
that are at each point parallel to the spin magnetization itself. 
Another way of seeing this is to notice that the ${xc}$ energy in LSDA is invariant not only with respect to global rotations of all the spins (as it should), 
but also with respect to local rotations which change the relative orientation of neighboring spins.  As a result, not only the global torque exerted by the ${xc}$ field,
but also the local torque vanishes, which implies that ${\vec{B}_{xc}}$ and the spin magnetization ${\vec{m}}$ are parallel.  
In order to introduce the physically expected dependence of the ${xc}$ energy on the relative orientation of neighboring spins
it is necessary to make the ${xc}$ functional dependent on the \emph{transverse} gradient of the spin magnetization.

One way to include the dependence of the ${xc}$ functional on  transverse gradients is to consider the spin-spiral-wave state of the uniform electron
gas.\cite{Overhauser:62,GiulianiVignaleSDW:05,KurthEich:09,EichGross:10} A perturbative correction to the LSDA based on the spin-spiral-wave state
has been obtained in Ref.~\onlinecite{KatsnelsonAntropov:03}. One of us has shown that it is possible to generalize the LSDA itself by
employing the spin-spiral-wave state as a reference system.\cite{EichGross:13}
The resulting approximation to ${E_{xc}}$ is not obtained through a gradient expansion, nevertheless
it introduces a dependence on the transverse gradients of the magnetization.  This approach does not deal with the longitudinal gradients,
since the latter vanish in the spin-spiral-wave state where the magnitude of the magnetization is constant.
In contrast to this, a standard gradient expansion  can in principle capture the effects of longitudinal gradients. 

A common approach to generalize existing collinear generalized gradient expansions (GGAs)
to non-collinear situations is to proceed by reinterpreting the standard collinear forms.~\cite{HobbsHafner:00a,HobbsHafner:00b,PeraltaFrisch:07}
GGAs can be written as
\begin{align}
  E_{xc}\!\bclr{n,\vec{m}} = \ind{^3r} F\!\blr{n_{+},n_{-},\gamma_{--},\gamma_{++},\gamma_{-+}} , \label{Exc_GGA}
\end{align}
where ${n_{+}\!\blr{\vr}}$ and ${n_{-}\!\blr{\vr}}$ are the densities of \emph{spin-up} and \emph{spin-down} electrons, respectively, and
\begin{align}
  \gamma_{\alpha \beta} & = \blr{\vD n_{\alpha}}\!\cdot\!\blr{\vD n_{\beta}} , \label{gamma}
\end{align}
with ${\alpha,\beta = \pm}$ . The function ${F}$ is taken from a GGA for collinear systems. A straightforward application to non-collinear
situations yields
\begin{subequations} \label{CNC}
  \begin{align}
    n_{\pm} & = \ahalf \blr{n \pm m} ~, \label{n_pm} \\
    m & = \sqrt{m_{x}^2+m_{y}^2+m_{z}^2} ~, \label{m} \\
    \gamma_{\pm\pm} & = \aquater \bigg( \blr{\vD n} \!\cdot\!\blr{\vD n} + \hat{\vec{m}} \!\circ\! \blr{\vD \vec{m}} \!\cdot\! \hat{\vec{m}} \!\circ\! \blr{\vD \vec{m}} \nn
      & \phantom{= \aquater \bigg(} {} \pm 2 \blr{\vD n} \!\cdot\! \hat{\vec{m}} \!\circ\! \blr{\vD \vec{m}} \bigg) ~, \label{gamma_aa} \\
    \gamma_{\pm\mp} & = \aquater \blr{ \blr{\vD n} \!\cdot\!\blr{\vD n} - \hat{\vec{m}} \!\circ\! \blr{\vD \vec{m}} \!\cdot\! \hat{\vec{m}} \!\circ\! \blr{\vD \vec{m}} } ~, \label{gamma_ab}
  \end{align}
\end{subequations}
where we introduced the notation ``${\cdot}$" for the scalar product in \emph{physical} space (contraction of ${\vD}$) and ``${\circ}$" for the scalar product
in \emph{spin} space (contraction of the components of ${\vec{m}}$). ${\hat{\vec{m}}}$ is the unit vector along ${\vec{m}}$. This implies that the local
direction of ${\vec{m}}$ is used to specify the \emph{up-down} direction, with  ${\hat{\vec{m}}}$ pointing, by definition, ``up". 

The main drawback  of this approximation is that 
{\em only} the longitudinal change of the magnetization is taken into account by projecting ${\vD\vec{m}}$ onto the local direction of the magnetization 
${\hat{\vec{m}}}$. As a consequence, the functional form is not sensitive to a change in the direction of the magnetization and
the ${xc}$ magnetic field ${\vec{B}_{xc}}$ is parallel to the spin magnetization. Therefore, 
${\vec{B}_{xc}}$ does not exert any local torque on ${\vec{m}}$ and the restoring force for, say, a spin wave, comes entirely from the
kinetic energy cost of rotating the spins from their equilibrium directions.

In a recent paper Scalmani and Frisch \cite{ScalmaniFrisch:12} proposed a modification
to make the ${xc}$ energy functional depend on the relative
orientation of the spins. This has been achieved by redefining the ${\gamma_{\alpha\beta}}$ that enter in the functional Eq.~\eqref{Exc_GGA}, i.e.~,
\begin{subequations} \label{SF}
  \begin{align}     
    \gamma_{\pm\pm} & = \aquater \bigg( \blr{\vD n} \!\cdot\!\blr{\vD n} + \blr{\vD \vec{m}} \!\cdot\!\circ\! \blr{\vD \vec{m}} \nn
      & \phantom{= \aquater \bigg(} {} \pm 2 f_{\vD} \sqrt{\blr{\vD n} \!\cdot\! \blr{\vD \vec{m}} \!\circ\! \blr{\vD n} \!\cdot\! \blr{\vD \vec{m}} } \bigg) ~, \label{gamma_aa_SF} \\
    \gamma_{\pm\mp} & = \aquater \blr{ \blr{\vD n} \!\cdot\!\blr{\vD n} -  \blr{\vD \vec{m}} \!\cdot\!\circ\! \blr{\vD \vec{m}} } ~, \label{gamma_ab_SF} \\
    f_{\vD} & = \sgn\!\bslr{\blr{\vD n}\!\cdot\!\hat{\vec{m}}\!\circ\!\blr{\vD \vec{m}}} ~. \label{f_SF}
  \end{align}
\end{subequations}
In order to reduce to the usual definition for collinear systems the sign of the third term in Eq.~\eqref{gamma_aa_SF} needs to be fixed by Eq.~\eqref{f_SF}.
At variance with definitions Eqs.~\eqref{gamma_aa}, \eqref{gamma_ab}, the functional, obtained employing definitions Eqs.\ \eqref{SF},
is sensitive to  variations of the direction  of the spin magnetization.
As a consequence ${\vec{B}_{xc}}$ is no longer aligned with the spin magnetization  and hence the ${xc}$ magnetic field exerts a torque on the magnetization.

On closer inspection, however, we observe that Eqs.~\eqref{SF} are too restrictive because the longitudinal and transverse gradients
of the magnetization are treated equally.
This can be seen by decomposing the total gradient of the magnetization into a \emph{longitudinal} component, parallel to the magnetization, and a \emph{transverse} component, 
perpendicular to it:
\begin{subequations} \label{grad_decomp}
  \begin{align}
    \vD \vec{m} & = \blr{\vD \vec{m}}_{\parallel}+\blr{\vD \vec{m}}_\perp ~, \label{grad_m_decomp} \\
    \blr{\vD \vec{m}}_{\parallel} & = \blr{\hat{\vec{m}} \!\circ\! \vD \vec{m}} \hat{\vec{m}} ~, \label{grad_m_longitudinal} \\
    \blr{ \vD \vec{m}}_{\perp} & = \blr{\hat{\vec{m}} \!\otimes\! \vD \vec{m}} \!\otimes\! \hat{\vec{m}} ~. \label{grad_m_transverse}
  \end{align}
\end{subequations}
Here  ``$\otimes$" represents the cross product with respect to spin indices.
Physically, the longitudinal component of the gradient describes the change of the magnitude of the magnetization, while the transverse component describes the change of its directions.
In Eqs.~\eqref{gamma_aa_SF}, \eqref{gamma_ab_SF} the longitudinal and transverse components of the gradient of the magnetization enter on equal footing as can be seen by writing,
\begin{align}
  \blr{\vD \vec{m}} \!\cdot\!\circ\! \blr{\vD \vec{m}} & \!=\! \blr{\vD \vec{m}}_{\parallel} \!\cdot\!\circ\! \blr{\vD \vec{m}}_{\parallel}
  \!+\! \blr{\vD \vec{m}}_{\perp} \!\!\cdot\!\circ\! \blr{\vD \vec{m}}_{\perp} ~, \nonumber
\end{align}
and
\begin{align}
  \blr{\vD n}\!\cdot\!\blr{\vD\vec{m}}\! & \circ\!\blr{\vD\vec{m}}\!\cdot\!\blr{\vD n} = \nn
  & \blr{\vD n}\!\cdot\!\blr{\vD\vec{m}}_{\parallel}\!\circ\!\blr{\vD\vec{m}}_{\parallel}\!\cdot\!\blr{\vD n} \nn
  + & \blr{\vD n}\!\cdot\!\blr{\vD\vec{m}}_{\perp}\!\circ\!\blr{\vD\vec{m}}_{\perp}\!\cdot\!\blr{\vD n} ~. \nonumber
\end{align}
This implies that, for example, starting from a state of spatially uniform magnetization, the ${xc}$ energy does not differentiate between a sinusoidal modulation of the magnitude
of the magnetization or a sinusoidal modulation of its direction -- as long as the amplitudes of the two modulations are the same.
The energy functional based on Eqs.~\eqref{SF} may be justified is when the magnitude of the magnetization tends to zero; for in that case there is
no preferred direction with respect to which one can assess whether the modulation is longitudinal or transverse.
For large ${m}$ (close to the saturation value), it should not be expected that a functional that was originally developed to  describe collinear spins would also
describe accurately variations of the direction of the magnetization. 

\begin{figure}
  \includegraphics{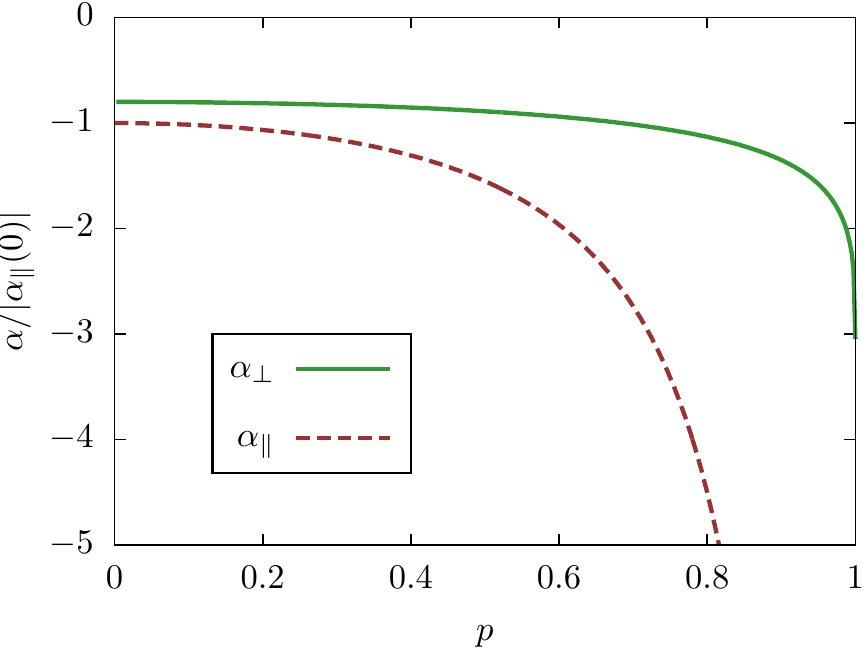}
  \caption{(Color online). The solid, green line shows the transverse coefficient $\alpha_\perp$ and the dashed, red line shows the longitudinal coefficient $\alpha_\parallel$.
    ${\alpha_{\perp}}$ and ${\alpha_{\parallel}}$ appear in the gradient expansion of the exchange energy Eq.~(\ref{Ex_GEA}).
    We address the {\em spurious} difference of the the two coefficients at vanishing spin polarization in Sec.\ (\ref{Gradients}).}
  \label{a1}
\end{figure}
In order to gain insight into these issues we will derive a gradient expansion for the exchange energy in the next section.
The correction to the local approximation to the exchange energy due to the gradient expansion reads,
\begin{align}
  & \delta E_{x}\bclr{n,\vec{m}} \approx \frac{1}{2} \ind{^3r} \Big( \alpha \abs{\vD n}^2 + \alpha_{\parallel} \abs{\hat{\vec{m}}\circ\blr{\vD \vec{m}}}^2  \nn
  & {} + 2 \alpha_{\times} \blr{\vD n}\!\cdot\!\blr{\hat{\vec{m}}\circ\blr{\vD \vec{m}}} 
  +  \alpha_{\perp}\abs{\hat{\vec{m}}\otimes\blr{\vD \vec{m}}}^2 \Big) . \label{Ex_GEA}
\end{align}
The main point we want to convey is the following: the two coefficients ${\alpha_{\parallel}}$ and ${\alpha_{\perp}}$, describing the
dependence of the ${xc}$ energy on the longitudinal and the transverse gradients respectively, differ. This is shown in FIG. \ref{a1} where we plot them
as functions of the  spin polarization ${p=m/n}$. One can clearly see that the difference between the two stiffnesses gets
bigger for larger spin magnetizations. Note that, a generic collinear GGA can reproduce ${\alpha}$, ${\alpha_{\parallel}}$ and ${\alpha_{\times}}$ in the
limit of small and slowly varying densities. Hence in our work we focus on obtaining the coefficient ${\alpha_{\perp}}$ and comparing it to ${\alpha_{\parallel}}$. 
 
\section{Derivation of the gradient expansion} \label{Gradients}

Consider a uniform electron gas of density ${n}$ on a rigid neutralizing background (jellium model). The electrons are polarized along an arbitrary direction (say ${z}$)  by a
uniform magnetic field. Let ${p}$ be the spin polarization, so that
\begin{align}
  p & = \frac{n_+-n_-}{n} \;\; , \;\; n = n_+ + n_- .
\end{align}
The equilibrium magnetization (in units of the Bohr magneton ${\mub}$) is
\begin{align}
  \vec{m} & = n p \hat{\vec{z}} = M \hat{\vec{z}} 
\end{align}
where ${\hat{\vec{z}}}$ is the unit vector in the ${z}$ direction. 
A small modulation of the charge density ${\delta n\!\blr{\vr}}$ (${\delta n \ll n}$) and the spin magnetization
\begin{align}
  \delta \vec{m}\!\blr{\vr} = \delta m_{\parallel}\!\blr{\vr} \hat{\vec{z}} + \delta \vec{m}_{\perp}\!\blr{\vr} ,
\end{align}
where ${\abs{\delta \vec{m}} \ll np}$ and ${\delta\vec{m}_{\perp}}$ is perpendicular to ${\hat{\vec{z}}}$, 
will change the energy of the system. The ${xc}$ energy associated with this modulation is
\begin{align}
  \delta E_{xc} = \ahalf \indotp{^3q}{3} \begin{pmatrix} \delta \tilde{n}\!\blr{\vq} & \delta\tilde{\vec{m}}\!\blr{\vq} \end{pmatrix} 
  \mat{I}\!\blr{\vq} \begin{pmatrix} \delta \tilde{n}\!\blr{\vq} & \delta\tilde{\vec{m}}\!\blr{\vq} \end{pmatrix}^{\mathrm{T}} , \label{DExc}
\end{align}
with the \emph{static} ${xc}$ kernel
\begin{align}
  \mat{I}\!\blr{\vq} & = \mat{\Pi}_{s}^{-1}\!\blr{\vq} - \mat{\Pi}^{-1}\!\blr{\vq} . \label{I_SDFT} 
\end{align}
Here ${\mat{\Pi}}$ and  ${\mat{\Pi}_{s}}$ are, respectively, the \emph{proper} response functions of the interacting and Kohn-Sham (non-interacting)
spin-polarized electron gas, \emph{with identical ground-state magnetizations}. ${\delta \tilde{n}\!\blr{\vq}}$ and ${\delta \tilde{\vec{m}}\!\blr{\vq}}$
are the Fourier transforms of ${\delta n\!\blr{\vr}}$ and ${\delta \vec{m}\!\blr{\vr}}$.
The \emph{proper} response function ${\mat{\Pi}}$, a ${4\times4}$-matrix in the case of non-collinear SDFT,
describes the change of the densities due to a perturbing potential,
\begin{align}
  \begin{pmatrix} \delta n\!\blr{\vq} \\ \delta m_{z}\!\blr{\vq} \\ \delta m_{x}\!\blr{\vq} \\ \delta m_{y}\!\blr{\vq} \end{pmatrix} & =
  \mat{\Pi}\!\blr{\vq}
  \begin{pmatrix} \delta v\!\blr{\vq} + \delta v_{\mathrm{H}}\!\blr{\vq} \\ \delta B_{z}\!\blr{\vq} \\ \delta B_{x}\!\blr{\vq} \\ \delta B_{y}\!\blr{\vq} \end{pmatrix} ~, \nonumber
\end{align}
where,
\begin{align}
  \mat{\Pi}\!\blr{\vq} & = \begin{pmatrix} \Pi_{nn}\!\blr{\vq} & \Pi_{nz}\!\blr{\vq} & 0 & 0 \\ \Pi_{zn}\!\blr{\vq} & \Pi_{zz}\!\blr{\vq} & 0 & 0 \\
    0 & 0 & \Pi_{xx}\!\blr{\vq} & 0 \\ 0 & 0 & 0 & \Pi_{yy}\!\blr{\vq} \end{pmatrix} ~. \label{Pi}
\end{align}
Note that in Eq.~\eqref{Pi} we use the fact that the unperturbed system, albeit being spin polarized, is a \emph{collinear} spin state.
The response function ${\mat{\Pi}}$ for any spin-polarized system decouples into two sectors, i.e.~, the density-longitudinal spin (${n-m_{\parallel}}$) sector
and the transverse spin (${m_{\perp}}$) sector. The two sectors are only coupled if the unperturbed system already exhibits a non-collinear spin magnetization. 
By symmetry ${\Pi_{xx}=\Pi_{yy}=\Pi_{\perp}}$ for a system polarized along the ${\hat{\vec{z}}}$-axis. 
Up to first-order in the interaction the difference between the inverse 
of the non-interacting response function  and that of the interacting response function is given by 
\begin{align}
  \bclr{\mat{\Pi}_{0}}^{-1}-\bclr{\mat{\Pi}}^{-1} \simeq
  \bclr{\mat{\Pi}_{0}}^{-1} \mat{\Pi}_{1} \bclr{\mat{\Pi}_{0}}^{-1} . \label{fx_static1}
\end{align}
The non-interacting response matrix ${\mat{\Pi}_{0}}$ is well known,\cite{GiulianiVignale:05} and the the ${n-m_{\parallel}}$ sector of the first-order interacting
response matrix ${ \mat{\Pi}_{1}}$ can be obtained from the calculation of Engel and Vosko.\cite{EngelVosko:90} Notice that both ${\mat{\Pi}_{0}}$ and ${\mat{\Pi}_{1}}$ 
in Eq.~(\ref{fx_static1}) are calculated in the presence of the same magnetic field ${B}$, which produces a ground-state magnetization $M$ in the interacting system.
In practice, the effect of this magnetic field is to produce a splitting ${\Delta=2 B}$ between the single-particle energies of spin-up and spin-down electrons.  
The static ${xc}$ kernel ${\mat{I}}$ (cf.~Eq.~\eqref{I_SDFT}), however, is given by the  difference between the inverse of the Kohn-Sham response function
and that of the interacting response function.
It is important to appreciate that ${\mat{\Pi}_{s}}$ and ${\mat{\Pi}_0}$ are not the same thing, even though they are both response functions of a non-interacting system.
The essential difference is that ${\mat{\Pi}_s}$ is calculated in the presence of the external field ${B_{s}}$ which produces the interacting magnetization ${M}$ in the non-interacting system,
without the assistance of electron-electron interactions, whereas ${\mat{\Pi}_{0}}$ is calculated in the field ${B}$, which produces ${M}$ in the interacting system 
but a different magnetization in a non-interacting system. To first order in the interaction the difference between ${\mat{\Pi}_{s}^{-1}}$ and ${\mat{\Pi}_{0}^{-1}}$ is given by
\begin{align}
  \bclr{\mat{\Pi}_{s}}^{-1} - \bclr{\mat{\Pi}_{0}}^{-1} \simeq
  \bclr{\mat{\Pi}_{s}}^{-1} \blr{ \partial_{\Delta} \mat{\Pi}_{0} }_{\Delta_{s}} 
  \bclr{\mat{\Pi}_{s}}^{-1} \Delta_{1} , \label{fx_static2}
\end{align}
with ${\Delta_{1}=-\tfrac{1}{\pi}\blr{\kfu-\kfd}}$ being the the change of the spin splitting to first order in the interaction.
Adding Eq.~(\ref{fx_static2}) to Eq.~(\ref{fx_static1}) we obtain the \emph{exchange} kernel 
\begin{align}
  \mat{I}_{1} & = \bclr{\mat{\Pi}_{s}}^{-1} \Blr{ \mat{\Pi}_{1} + \blr{\partial_{\Delta} \mat{\Pi}_{0}}_{\Delta_{s}} \Delta_{1} }\bclr{\mat{\Pi}_{s}}^{-1} \label{fx_static} ~,
\end{align}
where we replace ${\mat{\Pi}_{0}\!\blr{\vq} \to \mat{\Pi}_{s}\!\blr{\vq}}$ in Eq.~\eqref{fx_static1} since we are working only up to first-order.
Note that only the transverse sector of the response matrix depends on the spin splitting ${\Delta}$ and therefore the ``anomalous''  contribution, Eq.~\eqref{fx_static2},
vanishes in the ${n-m_{\parallel}}$ sector. From now on the shift ${\mat{\Pi}_{1}\to\mat{\Pi}_{1} + \blr{\partial_{\Delta} \mat{\Pi}_{0}}_{\Delta_{s}} \Delta_{1}}$
is implied when we refer to ${\mat{\Pi}_{1}}$ in the ${m_{\perp}}$ sector.

In order to make the connection to the gradient expansion we consider ${\delta \tilde{n}\!\blr{\vq}}$ and ${\delta \tilde{\vec{m}}\!\blr{\vq}}$ that
are significant only for small  ${q}$, i.e.~, we consider \emph{slow} modulations of the spin magnetization in space. This means that we can expand Eq.~\eqref{I_SDFT}
\begin{align}
  \mat{I}_{1}\!\blr{\vq} & \approx \mat{I}_{1}\!\blr{0} + \mat{\alpha}\!\blr{\frac{q}{\kf}}^2 ~, \label{I_SDFT_expand}
\end{align}
where
\begin{align}
  \mat{\alpha} & = \begin{pmatrix} \alpha_{\phantom{\parallel}} & \alpha_{\times} & 0 & 0 \\ 
    \alpha_{\times} & \alpha_{\parallel} & 0 & 0 \\ 
    0 & 0 & \alpha_{\perp} & 0 \\ 
    0 & 0 & 0 & \alpha_{\perp} \end{pmatrix} , \label{alpha}
\end{align}
is the ${q^2}$-coefficient of the static ${xc}$ kernel.
Since we are considering charge and spin \emph{modulations}, which implies that the ${q=0}$ component of ${\delta n}$ and ${\delta\vec{m}}$ vanishes,
transforming Eq.~\eqref{DExc} back to real space yields Eq.~\eqref{Ex_GEA}, i.e.\ , the correction due to gradients of the densities ${n}$ and ${\vec{m}}$.

\begin{figure}[hb]
  \begin{fmffile}{vertex}
    \begin{fmfgraph}(50,50)
      \fmfleft{i1,i2,i3}
      \fmfright{o1,o2,o3}
      \fmf{fermion,left=0.3,tension=1}{i2,v1,o2}
      \fmf{fermion,left=0.3,tension=1}{o2,v2,i2}
      \fmf{phantom,tension=5}{i3,v1,o3}
      \fmf{phantom,tension=5}{i1,v2,o1}
      \fmf{photon}{v1,v2}
      \fmfv{decor.shape=circle,decor.filled=empty,decor.size=5}{i2}
      \fmfv{decor.shape=circle,decor.filled=empty,decor.size=5}{o2}
      \fmfv{decor.shape=circle,decor.filled=empty}{i2}
      \fmfv{decor.shape=circle,decor.filled=empty}{o2}
    \end{fmfgraph}
  \end{fmffile} \hspace{0.1cm}
  \begin{fmffile}{selfenergy1}
    \begin{fmfgraph}(50,50)
      \fmfleft{i1,i2,i3}
      \fmfright{o1,o2,o3}
      \fmf{fermion,left=0.2,tension=1}{i2,v1,v2,o2}
      \fmf{fermion,left=0.6,tension=1}{o2,i2}
      \fmf{phantom,tension=5}{i3,v1}
      \fmf{phantom,tension=5}{v2,o3}
      \fmf{phantom,tension=1}{i1,v1}
      \fmf{phantom,tension=1}{v2,o1}
      \fmf{photon,right=0.5}{v1,v2}
      \fmfv{decor.shape=circle,decor.filled=empty,decor.size=5}{i2}
      \fmfv{decor.shape=circle,decor.filled=empty,decor.size=5}{o2}
    \end{fmfgraph}
  \end{fmffile} \hspace{0.1cm}
  \begin{fmffile}{selfenergy2}
    \begin{fmfgraph}(50,50)
      \fmfleft{i1,i2,i3}
      \fmfright{o1,o2,o3}
      \fmf{fermion,left=0.6,tension=1}{i2,o2}
      \fmf{fermion,left=0.2,tension=1}{o2,v1,v2,i2}
      \fmf{phantom,tension=5}{i1,v2}
      \fmf{phantom,tension=5}{v1,o1}
      \fmf{phantom,tension=1}{i3,v2}
      \fmf{phantom,tension=1}{v1,o3}
      \fmf{photon,right=0.5}{v1,v2}
      \fmfv{decor.shape=circle,decor.filled=empty,decor.size=5}{i2}
      \fmfv{decor.shape=circle,decor.filled=empty,decor.size=5}{o2}
    \end{fmfgraph}
  \end{fmffile}
  \caption{first-order diagrams for the proper response function. The left diagram is the \emph{vertex} contribution and the other two
    diagrams correspond to \emph{self-energy} insertions.} \label{1st_order_response}
\end{figure}
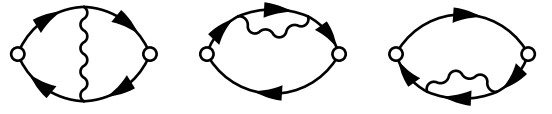
In order to determine the coefficients of the GEA we compute the first-order correction to the response matrix 
due to the \emph{vertex} and \emph{self-energy} diagrams depicted in FIG. \ref{1st_order_response}. In the static limit
the sum of these diagrams can be written in terms of the quantity,
\begin{widetext}
  \begin{align}
    C_{\sigma \sigma\pri}\!\blr{\vq} = -\otp{5}\ind{^3k_{1}}\ind{^3k_{2}} &
    \left(\frac{n_{\sigma}\!\blr{\vk_1}n_{\sigma}\!\blr{\vk_2} \blr{\blr{\vk_1-\vk_2}\cdot\vq}^2}
           {\blr{\vk_1-\vk_2}^2\blr{\vk_1\cdot\vq + \ahalf q^2 + \Delta_{\sigma\sigma\pri}}^2\blr{\vk_2\cdot\vq + \ahalf q^2+\Delta_{\sigma\sigma\pri}}^2} \right. \nn
           & \left. + \frac{n_{\sigma\pri}\!\blr{\vk_1}n_{\sigma\pri}\!\blr{\vk_2} \blr{\blr{\vk_1-\vk_2}\cdot\vq}^2}
           {\blr{\vk_1-\vk_2}^2\blr{\vk_1\cdot\vq + \ahalf q^2-\Delta_{\sigma\sigma\pri}}^2\blr{\vk_2\cdot\vq + \ahalf q^2-\Delta_{\sigma\sigma\pri}}^2} \right. \nn
           & \left. + \frac{2 n_{\sigma}\!\blr{\vk_1}n_{\sigma\pri}\!\blr{\vk_2} \blr{\blr{\vk_1+\vk_2+\vq}\cdot\vq}^2}
           {\blr{\vk_1+\vk_2+\vq}^2\blr{\vk_1\cdot\vq + \ahalf q^2+\Delta_{\sigma\sigma\pri}}^2\blr{\vk_2\cdot\vq + \ahalf q^2-\Delta_{\sigma\sigma\pri}}^2} \right) . \label{C}
  \end{align}
\end{widetext}
The spin splitting satisfies ${\Delta_{\sigma\sigma\pri}=-\Delta_{\sigma\pri\sigma}}$. The Kohn-Sham spin splitting is given by
${\bclr{\Delta_{s}}_{\sigma\sigma\pri} = \ahalf\blr{{\kf}_{\sigma}^2-{\kf}_{\sigma\pri}^2}}$.
We can write the components of the response matrix ${\mat{\Pi}_{1}}$ in terms of ${C_{\sigma\sigma\pri}}$, i.e.~,
\begin{subequations} \label{P1}
  \begin{align}
    \Pi_{1, \parallel} & = \Pi_{1, nn} = \Pi_{1, zz} = C_{\uparrow\uparrow} + C_{\downarrow\downarrow} ~, \label{P1_parallel} \\
    \Pi_{1, \times} & = \Pi_{1, nz} = \Pi_{1, zn} = C_{\uparrow\uparrow} - C_{\downarrow\downarrow} ~, \label{P1_cross} \\
    \Pi_{1, \perp} & = \Pi_{1, xx} = \Pi_{1, yy} = C_{\uparrow\downarrow} + C_{\downarrow\uparrow} + \bar{\Pi}_{1, \perp} ~. \label{P1_perp} \\
    \bar{\Pi}_{1,\perp} & = \blr{\partial_{\Delta} \Pi_{0, \perp}}_{\Delta_{s}} \Delta_{1} ~. \label{P1_anomalous}
  \end{align}
\end{subequations}
In Eq.\ \eqref{P1_anomalous} we defined the ``anomalous'' contribution ${\bar{\Pi}_{1,\perp}}$ arising due to the difference of ${\mat{\Pi}_{0}}$
and ${\mat{\Pi}_{s}}$, discussed previously.

\begin{figure}[t]
  \includegraphics{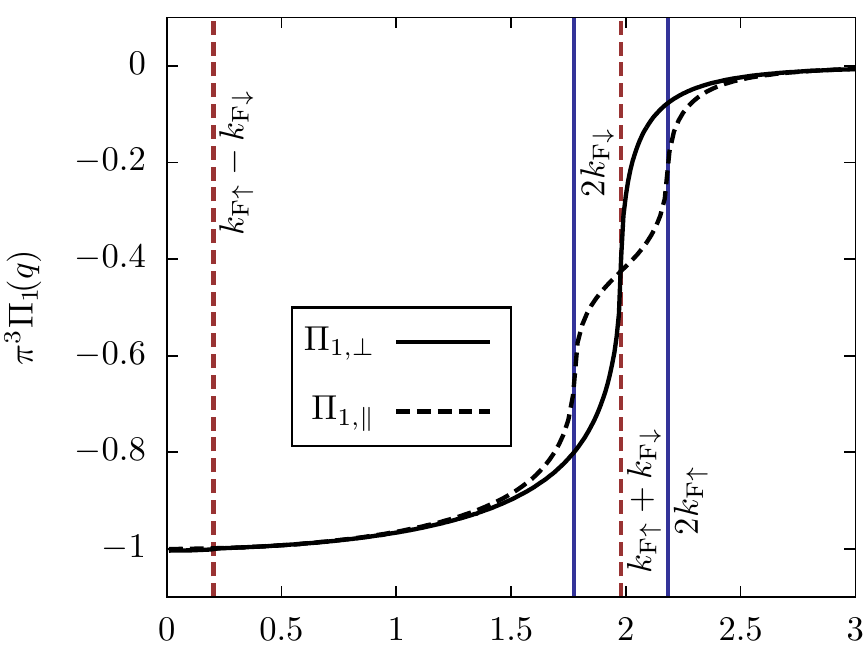}
  \includegraphics{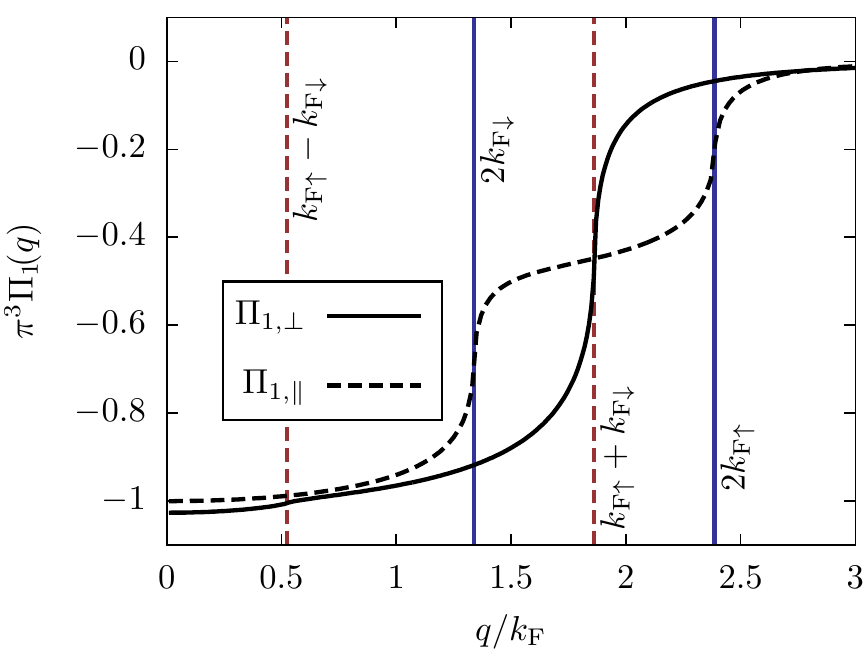}
  \caption{(Color online). Panels show the first-order correction to the longitudinal and transverse components of the response matrix. 
    The upper panel shows the longitudinal and the transverse response function
    for small spin polarization (${p=0.3}$). The lower panel shows the longitudinal and transverse response for large spin polarization (${p=0.7}$).
    The vertical solid, blue lines are at ${2\kfd}$ and ${2\kfu}$ and represent the important scales for the longitudinal response function ${\Pi_{\parallel}}$. 
    The vertical dashed, red lines are at ${\kfu-\kfd}$ and ${\kfu+\kfd}$ and are the relevant scales for the transverse response function ${\Pi_{\perp}}$.}
  \label{P1_0.3_0.7}
\end{figure}

\begin{figure}
  \includegraphics{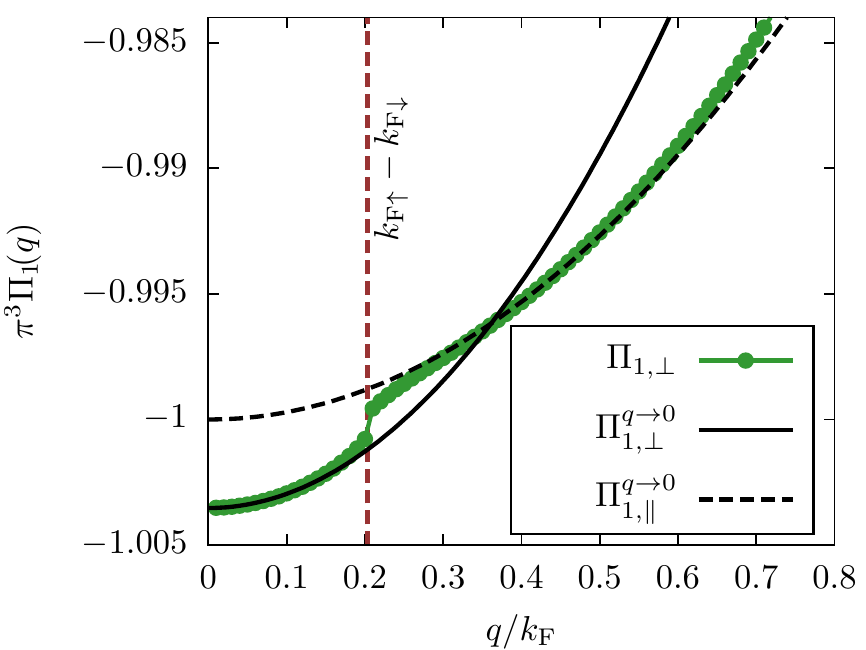}
  \caption{(Color online). Plot corresponds to the small-${q}$ region of the upper panel in FIG. \ref{P1_0.3_0.7}.
    The green line with dots shows the first-order correction of the transverse response matrix for a spin polarization of ${p=0.3}$.
    The vertical dashed, red line is at ${\kfu-\kfd}$. For comparison we show, as the solid line, the small-${q}$ expansion ${\Pi_{1,\perp}^{q\to 0}}$ of the transverse response function
    and, as the dashed line, the small-${q}$ expansion ${\Pi_{1,\parallel}^{q\to 0}}$ of the longitudinal response function.}
  \label{P1_0.3_smq}
\end{figure}

\begin{figure}
  \includegraphics{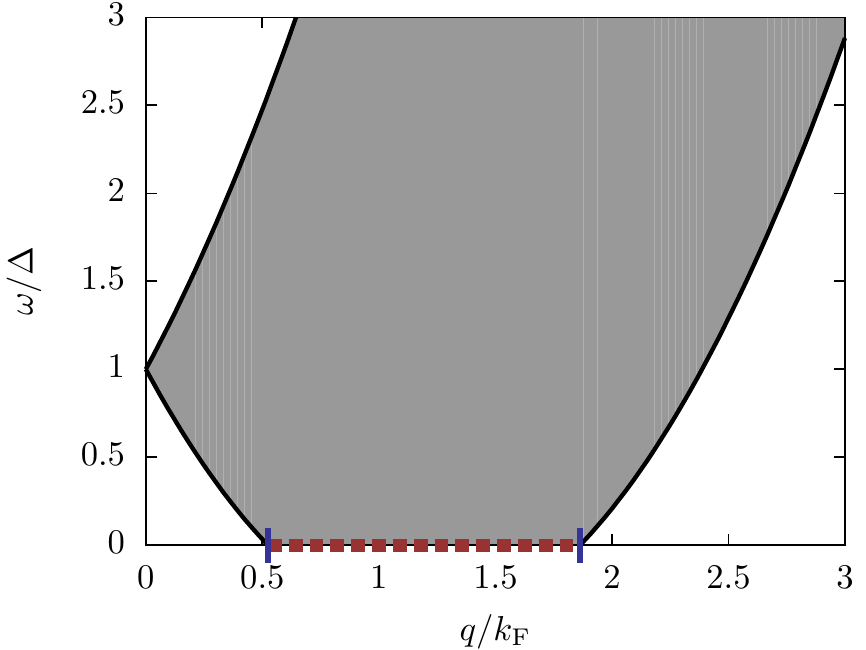}
  \caption{(Color online). Stoner continuum for spin polarization ${p=0.7}$. The shaded region represents the allowed spin-flip processes.
    At zero frequency the Stoner continuum lies between ${q=\kfu-\kfd}$ and ${q=\kfu+\kfd}$. This is depicted by the dashed, red line along the ${x}$-axis.}
  \label{stoner_continuum}
\end{figure}

In FIG. \ref{P1_0.3_0.7} we show the longitudinal component ${\Pi_{1, \parallel}}$ and the transverse component ${\Pi_{1, \perp}}$ of the response function
for small (${p=0.3}$) and large (${p=0.7}$) spin polarizations, respectively. ${\Pi_{1, \parallel}}$ exhibits the most structure at wave vectors ${q=2\kfu}$ and ${q=2\kfd}$
whereas ${\Pi_{1, \perp}}$ changes strongly around ${q=\kfu-\kfd}$ and ${q=\kfu+\kfd}$.
FIG. \ref{P1_0.3_0.7} shows clearly the difference between the longitudinal and the transverse response. The largest difference occurs in the region 
between ${2\kfd}$ and ${2\kfu}$ with a change in sign at ${\kfu+\kfd}$. This means that the  difference between the two response functions is bigger for larger spin polarizations ${p}$.
Moreover,  for the larger spin polarizations (lower panel of FIG. \ref{P1_0.3_0.7}) one can see that the longitudinal and transverse response functions also differ
in the region between ${0}$ and ${\kfu-\kfd}$.
In FIG. \ref{P1_0.3_smq} we show the transverse response function in this region for a spin polarization of ${p=0.3}$. There we also plot, for comparison, the \emph{small}-$q$
expansions for the longitudinal and transverse response function. 
The relevant scale of the wave-vector dependence for ${\Pi_{1, \parallel}}$ is set by
the onset of the particle-hole continuum in the spin-up (spin-down) channel, which lies between ${q=0}$ and ${q=2\kfu}$ (${q=2\kfd}$). 
For ${\Pi_{1, \perp}}$, however, the scale of the wave-vector dependence is set by the \emph{spin-flip} particle-hole continuum, 
the so-called Stoner continuum (cf.\ FIG.\ \ref{stoner_continuum}) which at zero frequency lies between ${q=\kfu-\kfd}$ and ${q=\kfu+\kfd}$.
This has important consequences for the expansion of the response matrix around ${q=0}$, because the components of the response function are
non-analytic when ${q}$ crosses into the region of the particle-hole or Stoner continuum. For the ${n-m_{\parallel}}$ sector we are always in
the region of the particle-hole continuum and the condition for ``small'' ${q}$ is given by ${q\ll 2\kfd}$. Note that this implies a vanishing
radius of convergence in the limit of ${p \to 1}$ for the small-${q}$ expansion in the ${n-m_{\parallel}}$ sector. This may be inferred
from FIG. \ref{P1_0.3_0.7} by recognizing that the steep change from ${\sim 1}$ to ${\sim 0.5}$ that occurs at ${q=2 \kfd}$ moves closer to ${q=0}$ for ${p \to 0}$ and hence
the  coefficient of ${q^2}$ (${\alpha_{\parallel}}$) diverges in this limit (cf.\ also FIG.\ \ref{a1}). For the ${m_{\perp}}$ sector, however, ``small'' ${q}$ means ${q\ll \kfu-\kfd}$, which
implies that ${q}$ is outside the region of the Stoner continuum if one expands around ${q=0}$. Opposite to the expansion in the ${n-m_{\parallel}}$ sector
the radius of convergence vanishes in the limit ${p \to 0}$. As we will see shortly this results in a mismatch of the ${q^2}$-coefficients  ${\alpha_{\parallel}}$
and ${\alpha_{\perp}}$ in the limit ${p \to 0}$. However, physically it is expected that the two coefficients coincide for ${p=0}$, because there is no preferred
direction to define the meaning of longitudinal versus transverse. The main point we want to make in this paper is that there is a  difference between the two coefficients
${\alpha_{\parallel}}$ and ${\alpha_{\perp}}$ at \emph{finite} polarization. This difference can be seen by comparing the small-${q}$ expansions for
the longitudinal and the transverse response shown in FIG. \ref{P1_0.3_smq}.

The small-${q}$ expansion for the ${n-m_{\parallel}}$ sector of the response matrix follows from the calculation of Engel and Vosko.\cite{EngelVosko:90} It reads
\begin{align}
  \Pi_{1, \parallel} & \approx - \frac{1}{\pi^3}\blr{1-\frac{\blr{1+p}^{2/3}+\blr{1-p}^{2/3}}{72\blr{1-p^2}^{2/3}} q^2} ~, \label{P1_parallel_small_q} \\
  \Pi_{1, \times} & \approx - \frac{1}{\pi^3} \frac{\blr{1+p}^{2/3}-\blr{1-p}^{2/3}}{72\blr{1-p^2}^{2/3}} q^2 ~. \label{P1_cross_small_q}
\end{align}
Together with the small-${q}$ expansion of the spin-resolved Lindhard function this yields the two ${q^2}$-coefficients ${\alpha_{\parallel}}$ and ${\alpha_{\times}}$
entering the gradient expansion Eq.\ \eqref{Ex_GEA},
\begin{subequations} \label{a1_nm_sector}
  \begin{align}
    \alpha_{\parallel} & = \frac{-5\pi}{72 \blr{3 \pi^2 n}^{4/3} } \;\;
    \frac{ Q\!\blr{p} + p R\!\blr{p} }{ \blr{1-p^2}^{4/3} } ~, \label{a1_parallel} \\
    \alpha_{\times} & = \frac{5\pi}{72 \blr{3 \pi^2 n}^{4/3} } \;\;
    \frac{ R\!\blr{p} + p Q\!\blr{p} }{ \blr{1-p^2}^{4/3} } ~, \label{a1_cross}
  \end{align}
\end{subequations}
where we defined,
\begin{subequations} \label{Q_def}
  \begin{align}
    Q\!\blr{p} & = \blr{1+p}^{1/3}+\blr{1-p}^{1/3} ~, \label{Q} \\
    R\!\blr{p} & = \blr{1+p}^{1/3}-\blr{1-p}^{1/3} ~. \label{R}
  \end{align}
\end{subequations}
In the limit of vanishing spin polarization (${p \to 0}$) this reduces to the well-known result,\cite{EngelVosko:90}
\begin{align}
  \alpha_{\parallel} & = -\frac{5\pi}{36 \blr{3 \pi^2 n}^{4/3}} , \label{a_parallel_m_0} \\
  \alpha_{\times} & = 0 . \label{a_cross_m_0}
\end{align}

Now we turn to the evaluation of the small-${q}$ expansion of the transverse response function. Due to the presence of the
spin splitting ${\Delta}$ in the denominators of Eq.~\eqref{C} it is possible to expand the integrand directly, provided ${q<\kfu-\kfd}$.
This simplifies the computation considerably compared to the calculation of the density-density response. 
The combination ${\tilde{\Pi}_{1, \perp}=C_{\uparrow\downarrow} + C_{\downarrow\uparrow}}$, which represents the contribution due to the diagrams depicted in FIG.\ \ref{1st_order_response},
yields after a straightforward calculation,
\begin{align}
  \tilde{\Pi}_{1, \perp} & \approx \frac{1}{\pi^3} \frac{4p^2}{27\Delta_s^4} \blr{\frac{q}{\kf}}^2  ~. \label{tP1_perp_q_expand}
\end{align}
From the transverse-spin Lindhard function we obtain the ``anomalous'' contribution,
\begin{align}
  \bar{\Pi}_{1, \perp} & \approx - \frac{R\!\blr{p}}{\pi^3}
  \blr{\frac{2 p}{3 \Delta_{s}^2}-\frac{10\Delta_{s}-3 S\!\blr{p}}{15\Delta_{s}^4} \blr{\frac{q}{\kf}}^2 } ~, \label{bP1_perp_q_expand}
\end{align}
with
\begin{align}
  S\!\blr{p} & = \blr{\blr{1+p}^{5/3}-\blr{1-p}^{5/3}} ~. \label{S}
\end{align}
Note that only the ``anomalous'' part contributes a constant (${q^0}$-coefficient) and that the ${q^2}$-coefficients for both partial contributions
diverge in the limit ${p \to 0}$, which can be seen by using the fact that ${\Delta_{s} \sim \tfrac{2}{3} p}$ for small ${p}$. 
However, they combine in the total transverse-spin response function,
\begin{align}
  \Pi_{1, \perp}\!\blr{q} & \approx - \frac{1}{\pi^3} \blr{\frac{ 8 p R\!\blr{p} }{ 3 T\!\blr{q} } + U\!\blr{p} \blr{\frac{q}{\kf}}^2} ~, \label{P1_perp_small_q}
\end{align}
where we defined
\begin{align}
  T\!\blr{p} & = \blr{\blr{1+p}^{2/3}-\blr{1-p}^{2/3}}^2 ~, \label{T} \\
  U\!\blr{p} & = \blr{ \frac{48 \blr{1-p^2}^{2/3} + 8 Q\!\blr{p} } {15 T\!\blr{p} Q^2\!\blr{p}} 
  - \frac{64 p^2}{27 T^2\!\blr{p}} } ~, \label{U}
\end{align}
in a way to yield a finite ${q^2}$-coefficient in the limit of vanishing spin magnetization, i.e.~,
\begin{align}
  \lim_{p \to 0} \Pi_{1, \perp}\!\blr{q} & \approx - \frac{1}{\pi^3}\blr{1 - \frac{1}{18} \blr{\frac{q}{\kf}}^2} ~. \label{P1_perp_q_expand_limit}
\end{align}
The ${q^0}$-coefficient reduces to the ${q^0}$-coefficient of the longitudinal-spin response function in this limit. The ${q^2}$-coefficient, however,
is \emph{twice} the ${q^2}$-coefficient of the longitudinal-spin response function for ${p \to 0}$. As mentioned earlier this has to be attributed to the fact that
the radius of convergence for the small-${q}$ expansion for the transverse-spin response function vanishes in that limit. Or, we can say that, for $p\to0$, $\Pi_\perp$ approaches
$\Pi_\parallel$ in a non-uniform manner, becoming increasingly close to the latter for ${q>\kfu-\kfd}$, but still retaining a finite difference in  slope at 
${q \simeq \kfu-\kfd}$ (cf.\ FIG.\ \ref{P1_0.3_smq}).  

Combining these results with the small-${q}$
expansion of the transverse-spin Lindhard function yields
\begin{align}
  \alpha_{\perp} = & \frac{-\pi}{\blr{3 \pi^2 n}^{4/3} }  \label{a_perp} \\
  & \times \frac{ 2 \blr{9 - p^2} - 9 \blr{1-p^2}^{1/3} \blr{Q\!\blr{p} + p R\!\blr{p}} }{ 15 p^2 T\!\blr{p} } ~, \nonumber
\end{align}
which reduces to ${\tfrac{4}{5}\alpha_{\parallel}}$ in the limit of vanishing spin polarization, due to the mismatch of the ${q^2}$-coefficients.

\section{Discussion} \label{Discussion}

The validity of the gradient expansion~(\ref{Ex_GEA})  requires requires that the wave vector $q$, which quantifies the rate of spatial variation of the spin magnetization,
 satisfy the inequality ${q<\kfu-\kfd}$. A sensible \emph{local}
measure for the wave vector of the transverse variation is ${q=\abs{\blr{\vD \vec{m}}_{\perp}}/m}$. Similarly the local ``spin-up'' and ``spin-down''
wave vectors may be defined locally by ${\kfu=\blr{3 \pi^2 n}^{1/3}\blr{1+\tfrac{m}{n}}}$ and ${\kfd=\blr{3 \pi^2 n}^{1/3}\blr{1-\tfrac{m}{n}}}$, respectively.
This suggest the criterion
\begin{align}
  & \frac{\abs{\blr{\vD \vec{m}}_{\perp}}}{m} < 2 \blr{3 \pi^2 n}^{1/3} \frac{m}{n} \nn
  \;\; \Leftrightarrow \;\; & \xi= \frac{\abs{\blr{\vD \vec{m}}_{\perp}} n}{2 \blr{3 \pi^2 n}^{1/3} m^2} < 1 , \label{gea_criterion}
\end{align}
for the validity of the GEA in real inhomogeneous systems.   The dimensionless transverse spin gradient ${\xi}$ plays a similar role as the dimensionless
density gradient ${s=\abs{\vD n}/\blr{ 2 \blr{3 \pi^2 n}^{1/3} n}}$. While ${s}$ characterizes whether a local density variations can be considered small
according to the condition ${q<2 \kf}$, ${\xi}$ determines whether a local transverse spin gradient can be considered sufficiently small.

Our analysis provides an exact limit that approximate ${xc}$ functionals for non-collinear SDFT should try to match.   It also provides useful indications on the best way to construct non-collinear functionals from existing  collinear ones.  For example, it shows that the approach suggested by Scalmani and Frisch applied to GGAs 
treats longitudinal and transverse spin-magnetization gradients in a restrictive fashion by including them on equal footing. This can only be justified in two limits, i.e.\ , the limit of weak polarization, or when the system is so strongly inhomogeneous that the local wave vector $q$ (defined above) exceeds ${2 \kfu}$. It appears that these conditions have been met in the cases computed so far.\cite{ScalmaniFrisch:12, BulikScuseria:13}

In general at finite polarization one should expect that the weights of the longitudinal and transverse gradients in Eqs.~(\ref{gamma_aa_SF}-\ref{f_SF}) should be different  and proportional to the coefficients $\alpha_\parallel$ and $\alpha_\perp$, at least as long as the exchange contribution is dominant and the local wave vector ${q}$ is below the local ${\kfu-\kfd}$. It remains a challenge to implement these ideas in a practically useful functional. We hope that the presented work will stimulate further development in the construction of functionals aimed at the description of non-collinear magnetic structures.

\begin{acknowledgments}
  We gratefully acknowledge support from DOE  Grant No. DE-FG02-05ER46203 (FGE, GV)
  and European Community's FP7, CRONOS project, Grant Agreement No. 280879 (SP).
\end{acknowledgments}

\begin{appendix}

\end{appendix}

\bibliography{SDFT_transverse_spin}

\end{document}